\documentstyle[aps,multicol,epsf,psfig]{revtex}

\newcommand{\beq}{\begin{equation}}
\newcommand{\eeq}{\end{equation}}
\newcommand{\beqa}{\begin{eqnarray}}
\newcommand{\eeqa}{\end{eqnarray}}

\def\opone{\leavevmode\hbox{\small1\kern-3.8pt\normalsize1}}

\begin{document}
\title{Universal cloning of continuous quantum variables}
\author{N. J. Cerf,$^{1,2}$ A. Ipe,$^1$ and X. Rottenberg$^1$}
\address{$^1$ Ecole Polytechnique, CP 165, Free University of Brussels,
1050 Brussels, Belgium\\
$^2$ Information and Computing Technologies Research Section,
Jet Propulsion Laboratory,\\
California Institute of Technology, Pasadena, CA 91109}

\date{October 1999}
\draft
\maketitle

\begin{abstract}

The cloning of quantum variables with continuous spectra
is analyzed. A universal---or Gaussian---quantum cloning machine is exhibited
that copies equally well the states of two conjugate variables 
such as position and momentum. 
It also duplicates all coherent states with a fidelity of 2/3. 
More generally, the copies 
are shown to obey a no-cloning Heisenberg-like uncertainty relation.

\end{abstract}


\begin{multicols}{2}

Most of the concepts of quantum computation
have been initially developed for discrete quantum variables,
in particular, binary quantum variables (quantum bits).
Recently, however, a lot of attention has been devoted to the study
of {\em continuous} quantum variables in informational or computation
processes, as they might be intrinsically easier to manipulate 
than their discrete counterparts. Variables with a continuous spectrum
such as the position of a particle or the amplitude of an
electromagnetic field have been shown to be useful to perform quantum 
teleportation\cite{bib_teleport}, quantum error correction\cite{bib_qec},
or, even more generally, quantum computation\cite{bib_lloyd_braun}.
Also, quantum cryptographic schemes relying on continuous
variables have been proposed\cite{bib_crypto}, while the concept
of entanglement  purification has been extended to
continuous variables\cite{bib_purif}. In this context,
a promising feature of quantum computation over continuous variables 
is that it can be performed in quantum optics
experiments by manipulating squeezed states
with {\em linear} optics elements such as beam splitters\cite{bib_braun}.

In this Letter, the problem of copying the state
of a system with continuous spectrum is investigated, and it is shown
that a particular unitary transformation, called cloning, can be found
that {\em universally} copies the position and momentum states.
Let us first state the problem in physical terms. Consider, as
an example of a continuous variable, the position $x$ of a particle 
in a one-dimensional space, and its canonically conjugate variable $p$.
If the wave function is a Dirac delta
function---the particle is fully localized in {\em position} space, then $x$
can be measured exactly, and several perfect copies of the
system can be prepared. However, such a cloning process fails
to exactly copy non-localized states, {\it e.g.},  momentum
states. Conversely, if the wave function is a plane wave with
momentum $p$---the particle is localized in {\em momentum} space, then 
$p$ can be measured exactly and one can again prepare perfect copies 
of this plane wave. However, such a ``plane-wave cloner'' is then
unable to copy position states exactly. In short, it is impossible
to copy perfectly the states of two conjugate variables such as
position and momentum or the quadrature amplitudes of an electromagnetic
field. This simply illustrates the famous no-cloning 
theorem\cite{bib_nocloning} for continuous variables.

In what follows, it is shown that a unitary {\em cloning} transformation
can nevertheless be found that provides two copies of a
system with a continuous spectrum, but at the price
of a non-unity cloning fidelity.
More generally, we show that a class of cloning machines 
can be defined that yield two imperfect copies of a continuous
variable, say $x$. The quality of the two copies
obey a no-cloning uncertainty relation akin to the Heisenberg
relation, implying that the product of the $x$-error variance on the 
first copy times the $p$-error variance on the second one
remains bounded by $(\hbar/2)^2$---it cannot be zero.
Within this class, a {\em universal} cloner can be found that provides 
two identical copies of a continuous system with the same error distribution 
for position {\em and} momentum states. 
This cloner effects Gaussian-distributed
position- and momentum-errors on the input variable, and
is the continuous counterpart of the universal cloner 
for quantum bits\cite{bib_ucm}. More generally, it duplicates
in a same manner the eigenstates of linear combinations
of ${\hat x}$ and ${\hat p}$, such as Gaussian wave packets 
or coherent states. The latter states are shown
to be cloned with a fidelity that is equal to 2/3.

In the following, we shall work in position basis, whose states
$|x\rangle$ are normalized according to $\langle x|x'\rangle=\delta(x-x')$.
We assume $\hbar=1$, so that the momentum eigenstates
are given by $|p\rangle =(2\pi)^{-1/2}\int dx \, {\rm e}^{ipx} |x\rangle $.
Let us define the maximally-entangled states
of two continuous variables,
\beq   \label{eq_EPR}
|\psi(x,p)\rangle = {1\over\sqrt{2\pi}} \int_{-\infty}^{\infty}
dx'\; {\rm e}^{ipx'} \; |x'\rangle_1 |x'+x\rangle_2
\eeq
where 1 and 2 denote the two variables, while 
$x$ and $p$ are two real parameters. Equation (\ref{eq_EPR}) 
is akin to the original Einstein-Podolsky-Rosen (EPR) 
state\cite{bib_EPR}, but parametrized
by the center-of-mass position and momentum.
It is easy to check that $|\psi(x,p)\rangle$ 
is maximally entangled as 
${\rm Tr}_1 \left( |\psi\rangle \langle\psi| \right)
= {\rm Tr}_2 \left( |\psi\rangle \langle\psi| \right)
= \openone / (2\pi)$ for all values of $x$ and $p$,
with ${\rm Tr}_{1,2}$ denoting partial traces with respect to variable
1 and 2, respectively. The states $|\psi\rangle$ 
are orthonormal, i.e., $\langle\psi(x',p')|\psi(x,p)\rangle=
\delta(x-x')\, \delta(p-p')$, and satisfy a closure relation 
\beq
\int\!\!\int_{-\infty}^{\infty} dx \, dp \; |\psi(x,p)\rangle
\langle\psi(x,p)| = \openone_1 \otimes \openone_2
\eeq
so they form an orthonormal basis of the joint Hilbert space
of variables 1 and 2. Interestingly, applying some unitary operator
on {\em one} of these two variables makes it possible to transform the
EPR states into each other. Specifically, let us define
a set of {\em displacement} operators ${\hat D}$ 
parametrized by $x$ and $p$,
\beq
{\hat D}(x,p) = {\rm e}^{-ix{\hat p}} \; {\rm e}^{ip{\hat x}} =
\int_{-\infty}^{\infty} dx' \; {\rm e}^{ipx'} \;
|x'+x\rangle\langle x'| 
\eeq
which form a continuous Heisenberg group. Physically, ${\hat D}(x,p)$ 
denotes a momentum shift of $p$ followed by a position shift of $x$. 
If ${\hat D}(x,p)$ acts on one of two entangled continuous variables,
it is straightforward to check that
\beq  \label{eq_shift}
\openone \otimes {\hat D}(x,p) \; |\psi(0,0)\rangle = |\psi(x,p)\rangle
\eeq
This will be useful to specify the errors induced by the continuous
cloning machines considered later on. Assume that the input variable
of a cloner is initially entangled with another (so-called reference) 
variable, so that their joint state is $|\psi(0,0)\rangle$. If
cloning induces, say, a position-shift error of $x$
on the copy, then the joint state of the reference and
copy variables will be $|\psi(x,0)\rangle$ as a result of
Eq.~(\ref{eq_shift}). Similarly, a momentum-shift error of $p$
will result in $|\psi(0,p)\rangle$. More generally, if
these $x$ and $p$ errors are distributed at random according
to the probability $P(x,p)$, then the joint state will be
the mixture 
\beq   \label{eq_mixture}
\int\!\!\int_{-\infty}^{\infty} dx \, dp \; P(x,p) \;
|\psi(x,p)\rangle\langle\psi(x,p)|
\eeq

Let us now consider a cloning machine defined as the unitary
transformation ${\hat{\cal U}}$ acting on three continuous
variables: the input variable (variable 2) supplemented with two auxiliary
variables, the blank copy (variable 3) and an ancilla (variable 4).
After applying ${\hat{\cal U}}$,
variables 2 and 3 are taken as the two outputs of the cloner, while
variable 4 (the ancilla) is simply traced over. 
Assume now that variable 1 denotes a reference variable that is 
initially entangled with the cloner input---their
joint state is $|\psi(0,0)\rangle_{1,2}$, while the auxiliary variables
3 and 4 are initially prepared in the entangled state
\beq   \label{eq_chi}
|\chi\rangle_{3,4} =
\int\!\!\int_{-\infty}^{\infty} dx \, dp \; f(x,p) \;
|\psi(x,-p)\rangle_{3,4}
\eeq
where $f(x,p)$ is an (arbitrary) complex amplitude function of $x$ and $p$.
The cloning transformation is defined as
\beq   \label{eq_defU}
{\hat{\cal U}}_{2,3,4} = {\rm e}^{-i({\hat x_4}-{\hat x_3}){\hat p_2} } \;
{\rm e}^{-i{\hat x_2}({\hat p_3}+{\hat p_4}) } 
\eeq
where ${\hat x_k}$ (${\hat p_k}$) is the position (momentum) operator 
for variable $k$. 
Then, the joint state of the reference, the two copies,
and the ancilla after cloning
is given by $|\Phi\rangle_{1,2,3,4}= 
\openone_1\otimes {\hat {\cal U}}_{2,3,4} \;
|\psi(0,0)\rangle_{1,2} \; |\chi\rangle_{3,4}$.
Using Eqs.~(\ref{eq_chi}) and (\ref{eq_defU}), it can be written as
\beq   \label{eq_phi}
|\Phi\rangle = 
\int\!\!\int_{-\infty}^{\infty} dx \, dp \; f(x,p) \;
|\psi(x,p)\rangle_{1,2} \; |\psi(x,-p)\rangle_{3,4}
\eeq
This is a very peculiar 4-variable state in that it can be reexpressed
in a similar form by exchanging variables 2 and~3, namely
\beq   \label {eq_phibis}
|\Phi\rangle = 
\int\!\!\int_{-\infty}^{\infty} dx \, dp \; g(x,p) \;
|\psi(x,p)\rangle_{1,3} \; |\psi(x,-p)\rangle_{2,4}
\eeq
with
\beq    \label{eq_dft}
g(x,p)= {1\over 2\pi}
\int\!\!\int_{-\infty}^{\infty} dx' \, dp' \; 
{\rm e}^{i(px'-xp')} \; f(x',p')
\eeq

Thus, interchanging the two cloner outputs amounts to substitute the function
$f$ with its two-dimensional Fourier transform $g$\cite{fn0}. 
This property is crucial 
as it ensures that the two copies suffer from {\em complementary}
position and momentum errors. 
Indeed, using Eq.~(\ref{eq_phi}) and tracing over variables 3 and 4,
we see that the joint state of the reference and the first output 
is given by Eq.~(\ref{eq_mixture}), with $|f|^2$ playing the role of $P$.
Hence, the first copy (called copy $a$ later on) 
is imperfect in the sense that the input variable gets
a random position- and momentum-shift error
drawn from the probability distribution $P_a(x,p)=|f(x,p)|^2$.
Similarly, tracing the state~(\ref{eq_phibis}) over variables 2 and 4  
implies that the second copy (or copy $b$) is affected by a position- and 
momentum-shift error distributed as $P_b(x,p)=|g(x,p)|^2$.
The complementarity between the quality of the two copies originates 
from the relation between the amplitude functions $f$ and $g$, 
i. e., Eq.~(\ref{eq_dft}), in close analogy with what was shown for
discrete quantum cloners\cite{bib_cerf}.

Now, let us apply the cloning transformation ${\hat {\cal U}}$ 
on an input {\em position} state $|x_0\rangle$.
We simply need to project the reference
variable onto state $|x_0\rangle$.
Applied to the initial joint state of the reference and the input
$|\psi(0,0)\rangle_{1,2}$, this projection operator
$|x_0\rangle\langle x_0|\otimes \openone$
yields $|x_0\rangle_1 |x_0\rangle_2$ up to a
normalization, so the input is indeed projected onto the desired state.
Applying this projector to the state $|\Phi\rangle$
as given by Eq.~(\ref{eq_phi}) 
results in the state
\beq
\int\!\!\int_{-\infty}^{\infty} dx \, dp \; f(x,p) \;
{\rm e}^{i p x_0} |x_0+x\rangle_2 \; |\psi(x,-p)\rangle_{3,4}
\eeq
for the remaining variables 2, 3, and 4.
The state of copy $a$ (or variable 2) is then obtained by
tracing over variables 3 and 4,
\beq  \label{eq_rho_a}
\rho_a=\int_{-\infty}^{\infty} dx \; P_a(x) \;
|x_0+x \rangle\langle x_0+x|
\eeq
where $P_a(x)=\int_{-\infty}^{\infty} dp \; P_a(x,p)$
is the position-error (marginal) distribution affecting copy $a$.
Hence, the first copy undergoes a random position error which
is distributed as $P_a(x)$.
Similarly, applying the projector on the alternate expression for 
$|\Phi\rangle$, Eq.~(\ref{eq_phibis}), and tracing over variables 2
and 4 results in a state $\rho_b$ of the second copy  
that is akin to Eq.~(\ref{eq_rho_a})
with $P_b(x)=\int_{-\infty}^{\infty} dp \; P_b(x,p)$.
The result of cloning an input {\em momentum} state
$|p_0\rangle$ can also be easily determined 
by applying a projector onto $|-p_0\rangle$
to the reference variable, so that the initial joint state of the reference
and the input is projected on $|-p_0\rangle_1 |p_0\rangle_2$.
Using Eqs. (\ref{eq_phi}) and (\ref{eq_phibis}), 
we obtain the analogous expressions for the state of copies $a$ and $b$,
\beq
\rho_{a(b)} = \int_{-\infty}^{\infty} dp \; P_{a(b)}(p) \;
|p_0+p \rangle\langle p_0+p| 
\eeq
where $P_{a(b)}(p)=\int_{-\infty}^{\infty} dx \; P_{a(b)}(x,p)$.
Consequently, the two copies undergo a random momentum error
distributed as $P_{a(b)}(p)$. The tradeoff between the quality 
of the copies can be expressed  by relating the
variances of the distributions $P_a(x,p)$ and $P_b(x,p)$.

Let us analyze this no-cloning complementary by applying
the Heisenberg uncertainty relation to the state
\beqa
|\zeta\rangle_{1,2} 
&=& \int\!\!\int_{-\infty}^{\infty} 
 dx\, dp\; f(x,p)   \; |x\rangle_1 |p\rangle_2 \nonumber\\
&=& \int\!\!\int_{-\infty}^{\infty}
 dx\, dp\; g(-x,-p) \; |p\rangle_1 |x\rangle_2
\eeqa
where $|p\rangle_{1(2)}$ denote the momentum states 
of the first (second) variable.
The two pairs of conjugate operators $({\hat x}_1,{\hat p}_1)$
and $({\hat p}_2,{\hat x}_2)$ give rise, respectively, to the 
two no-cloning uncertainty relations
\beqa   \label{eq_uncert}
(\Delta x_a)^2 (\Delta p_b)^2 &\ge& 1/4  \nonumber\\
(\Delta x_b)^2 (\Delta p_a)^2 &\ge& 1/4
\eeqa
where $(\Delta x_a)^2$ and $(\Delta x_b)^2$
denote the variance of $P_a(x)$ and $P_b(x)$, respectively.
(The analogous notation holds for the momentum-shift distributions
affecting both copies.) Consequently, if the cloning process
induces a small position (momentum) error on the first copy, 
then the second copy is necessarily affected by a large 
momentum (position) error.

We now focus our attention on a symmetric and universal
continuous cloner that saturates the above no-cloning uncertainty
relations. We restrict ourselves to solutions of the
form $f(x,p)=q(x)\, Q(-p)$ where
$Q(p)=(2\pi)^{-1/2}\int_{-\infty}^{\infty} dx\; 
{\rm e}^{-ipx} \; q(x)$ is the Fourier transform of $q(x)$.
It can be checked that this choice satisfies the symmetry requirement,
$g(x,p)=f(x,p)$ \cite{fn1}. 
Now, for the cloner to act equally on position
and momentum states, $q(x)$ must be equal to its Fourier transform.
Hence, the universal continuous cloner corresponds to the choice
\beq 
f(x,p)={1\over \sqrt{\pi}} \; {\rm e}^{- {x^2+p^2 \over 2} }
\eeq
so that $P_{a(b)}(x,p) = {\rm e}^{- (x^2+p^2) }/\pi$
is simply a bi-variate Gaussian of variance 1/2 on $x$- and $p$-axis.
The two auxiliary variables must then be prepared in the state
\beq  \label{eq_vacuum}
|\chi\rangle = {1\over\sqrt{\pi}}
\int\!\!\int_{-\infty}^{\infty} dy\, dz \; {\rm e}^{-(y^2+z^2)/2} 
\; |y\rangle  \; |y+z\rangle
\eeq
The resulting transformation effected by this universal cloner
on an input position state $|x\rangle$ is given by
\beqa  \label{eq_U}
\lefteqn{ |x\rangle\;|0\rangle\;|0\rangle \to {1\over\sqrt{\pi}}
\int\!\!\int_{-\infty}^{\infty} dy\, dz \; {\rm e}^{-(y^2+z^2)/2} } 
 \hspace{3 truecm} \nonumber\\
& & \times \; |x+y\rangle \; |x+z\rangle \; |x+y+z\rangle
\eeqa
where the three variables denote the two copies and the ancilla,
respectively. It is easy to check that Eq.~(\ref{eq_U})
implies Eq.~(\ref{eq_rho_a}) and  its counterpart for copy $b$
with $P_a(x)=P_b(x)=\exp(-x^2)/\sqrt{\pi}$, so that both
copies are affected by a Gaussian-distributed position error
of variance 1/2.
The choice $(\Delta x_a)^2 = (\Delta x_b)^2 = (\Delta p_a)^2 = (\Delta p_b)^2
=1/2$ ensures that the cloner is {\em universal},
that is, position {\em and} momentum states are copied
with the same error variance. The value 1/2 implies that the
cloner is optimal among the class of cloners considered here
in view of Eq.~(\ref{eq_uncert}). Furthermore,
the cylindric symmetry of $f(x,p)$ [i.e., it depends only
on the radial coordinate $(x^2+p^2)^{1/2}$] implies that this
cloner copies the eigenstates of any operator 
of the form $c\,{\hat x}+d\,{\hat p}$
with the same error distribution, as we will show.

Let us first determine the operation of this universal cloner on an
arbitrary state $|\xi\rangle$ expressed in position basis as
$\int_{-\infty}^{\infty} dx \; \xi(x) \; |x\rangle$.
For this, we project the reference variable onto state
$|\xi^*\rangle$, i.e., the
state obtained by changing $\xi(x)$ into its complex conjugate.
This projector $|\xi^*\rangle\langle\xi^*|\otimes \openone$
applied on the initial state $|\psi(0,0)\rangle_{1,2}$
yields the state $|\xi^*\rangle_1 |\xi\rangle_2$ up to a
normalization, so the input is indeed projected onto $|\xi\rangle$.
Now, applying this projector to the state
$|\Phi\rangle$ after cloning implies that the three remaining
variables are projected onto the state
\beq
\int\!\!\int_{-\infty}^{\infty} dx \, dp \; f(x,p) \;
|\xi(x,p)\rangle_2 \; |\psi(x,-p)\rangle_{3,4}
\eeq
where
$ |\xi(x,p)\rangle = {\hat D}(x,p) \;  |\xi\rangle =
\int_{-\infty}^{\infty} dx' \; \xi(x') {\rm e}^{ipx'}   
|x'+x\rangle $
is the input state $|\xi\rangle$ affected
by a momentum shift of $p$ followed by a position shift of $x$.
This yields
\beq  \label{eq_general}
\rho_{a(b)} = \int\!\!\int_{-\infty}^{\infty} dx \; dp \; P_{a(b)}(x,p) \;
|\xi(x,p)\rangle\langle \xi(x,p)| 
\eeq
so that the two outputs are mixtures of the $|\xi(x,p)\rangle$ states,
with $x$ and $p$ distributed according to $P_{a(b)}(x,p)$.
Expressed in terms of Wigner distributions, Eq.~(\ref{eq_general})
implies that $W_{\rm out}(x,p) = W_{in}(x,p) \circ P(x,p)$ with
$\circ$ denoting convolution. In particular, the universal cloner 
simply effects a spreading out of the input Wigner function
by a bi-variate Gaussian of variance 1/2.

These considerations can be easily generalized to any pair of
canonically conjugate variables in a rotated phase space. 
First note that, using the Baker-Hausdorff formula
and $[{\hat x},{\hat p}]=i$, the displacement operator
can be rewritten as 
${\hat D}(x,p)={\rm e}^{-ixp/2}\; {\rm e}^{i(p{\hat x}-x{\hat p})}$.
Consider now any pair of observables ${\hat u}$ and
${\hat v}$ satisfying the commutation rule $[{\hat u},{\hat v}]=i$. 
Let ${\hat u} = c\, {\hat x}+d\, {\hat p}$ and
${\hat v} = -d\, {\hat x}+c\, {\hat p}$,
where $c$ and $d$ are real and satisfy $c^2+d^2=1$. 
It is easy to check that
${v{\hat u}-u{\hat v}}={p{\hat x}-x{\hat p}}$, where the variables
$u$ and $v$ are defined just as ${\hat u}$ and
${\hat v}$, so that ${\hat D}$ takes a similar form in terms of
${\hat u}$ and ${\hat v}$ (up to an irrelevant phase).
Therefore, as a consequence of 
the cylindric symmetry of the Gaussian $|f(x,p)|^2$, the
eigenstates $|u\rangle$ of the observable ${\hat u}$ 
undergo a random shift of $u$ that is distributed as
$\exp(- u^2 )/\sqrt{\pi}$. (The position and momentum states
are just two special cases of this.)
We can also treat the cloning of coherent states (Gaussian wave packets)
by considering the complex rotation that defines the annihilation and
creation operators ${\hat a}=({\hat x}+i{\hat p})/\sqrt{2}$
and ${\hat a}^{\dagger}=({\hat x}-i{\hat p})/\sqrt{2}$.
The displacement operator can then be written (up to an irrelevant
phase) in the usual form
${\hat D}(\alpha)={\rm e}^{\alpha {\hat a}^{\dagger}-\alpha^*{\hat a} }$,
where $\alpha=(x+ip)/\sqrt{2}$ is a c-number
that characterizes the position and momentum shift.
This operator transforms
the coherent state $|\alpha_0\rangle$ (i.e., the eigenstate of 
${\hat a}$ with eigenvalue $\alpha_0$) into
${\hat D}(\alpha)\, |\alpha_0\rangle = {\rm e}^{i\theta}
|\alpha_0+\alpha\rangle$, where $\theta={\rm Im}(\alpha \, \alpha_0^*)$.
Thus, if the input of the universal cloner is
a coherent state $|\alpha_0\rangle$, its two outputs are
a mixture of coherent states
characterized by the density operator
\beq   \label{eq_rho_coherent}
\rho=\int d^2\alpha \; G(\alpha) \; 
|\alpha_0+\alpha\rangle\langle\alpha_0+\alpha|
\eeq
where the integral is over the complex plane, and
$G(\alpha)=2 \exp(-2|\alpha|^2)/\pi$ is a Gaussian distribution
in $\alpha$ space. Then, using $|\langle \alpha | \alpha' \rangle|^2
= \exp(-|\alpha-\alpha'|^2)$,
it is easy to calculate the cloning fidelity:
\beq
f=\langle\alpha_0|\rho|\alpha_0\rangle
={2\over\pi} \int d^2\alpha \; {\rm e}^{-3|\alpha|^2} 
={2\over 3}
\eeq
This fidelity does not depend on $\alpha_0$, so 
it is universal for {\em all} coherent states.

Finally, we consider the cloning of quadrature squeezed states, defined as
the eigenstates of ${\hat b}=({\hat x}/\sigma+i\sigma{\hat p})/\sqrt{2}$,
where $\sigma$ is a real parameter. These states can be denoted
as $|\beta\rangle$, where $\beta=(x/\sigma+i\sigma p)/\sqrt{2}$ is a c-number.
We have again 
${\hat D}(\beta)={\rm e}^{\beta {\hat b}^{\dagger}-\beta^*{\hat b} }$,
so that ${\hat D}(\beta)|\beta_0\rangle=|\beta_0+\beta\rangle$
up to a phase. 
In order to keep the fidelity maximum, however, we must use here
a {\em non-universal} cloner defined by
\beq 
f(x,p)={1\over \sqrt{\pi}} \; {\rm e}^{- {x^2 \over 2 \sigma^2}+{\sigma^2 p^2
\over 2} }
\eeq
Both copies yielded by this cloner are affected by an $x$-error
of variance $\sigma^2/2$ and a $p$-error variance of $1/(2\sigma^2)$,
which implies that the density operator has the same form as
Eq.~(\ref{eq_rho_coherent}) with $G(\beta)=2\exp(-2|\beta|^2)/\pi$.
As a consequence, 
there exists a specific cloning machine for each value of $\sigma$
that copies all squeezed states corresponding to that $\sigma$ 
with a fidelity of 2/3. In contrast, cloning these states using the
universal cloner gives a fidelity that decreases as squeezing
increases.

We have shown that a universal cloning machine 
for continuous quantum variables can be defined that transforms
position (momentum) states into a Gaussian-distributed 
mixture of position (momentum) states with an error variance of 1/2.
It is universal, as the eigenstates of any linear combination
of ${\hat x}$ and ${\hat p}$ are copied with the same error distribution.
In particular, it duplicates all coherent states with a fidelity of 2/3.
We conjecture that this cloning fidelity is optimal.
An experimental realization of this universal cloner
could be envisaged based on the manipulation of modes
of the electromagnetic field. The cloning transformation ${\hat{\cal U}}$ 
would then couple two auxiliary modes to the input mode to be copied.
Since ${\hat{\cal U}}$ amounts to a sequence 
of ``continuous controlled-{\sc not} gates'', it could be
implemented by pairwise optical QND coupling 
between these three modes\cite{bib_braun}. 
As a final remark, it is worth
noting that the two auxiliary modes must be prepared in
state (\ref{eq_vacuum}), which is simply the product vacuum state 
$|0\rangle_3 |0\rangle_4$ processed by a controlled-{\sc not} gate 
${\rm e}^{-i{\hat x_3}{\hat p_4}}$. This suggests that the noise
that inevitably arises in the cloning of the input mode is simply linked to
the vacuum fluctuations of the two auxiliary modes.


This work was supported in part by DARPA/ARO under 
grant \# DAAH04-96-1-3086 through the QUIC Program.
N.~J.~C. is grateful to Samuel Braunstein, Jonathan Dowling, and 
Serge Massar for very helpful comments on this manuscript.

\end{multicols}

\begin{references}
\bibitem{fn0} More precisely,
$x'\to p$  is an inverse Fourier transform while $p' \to x$ is a
direct Fourier transform.
\bibitem{fn1} This solution is not unique as
symmetry only requires in general that $|g(x,p)|^2=|f(x,p)|^2$.
\end{references}
\end{document}